\documentclass[]{cupconf}
\usepackage{graphicx,psfig}

\title[Circumnuclear star forming regions]
      {The metallicity of circumnuclear star forming regions}
      
\author[A.~I.~D\'\i az et al.]{\'Angeles I. D\'\i az $^1$ 
\\ Elena Terlevich$^2$
\\ Marcelo Castellanos$^{1,3}$
\\ Guillermo H\"agele$^1$} 
\affiliation{$^1$ Universidad Aut\'onoma de Madrid, Spain \\ $^2$ Instituto Nacional de Astrof\'\i sica, \'Optica y Electr\'onica, Puebla, M\'exico \\ $^3$Instituto de Estructura de la Materia, CSIC, Madrid, Spain}   

\begin{document}
\maketitle

\begin{abstract}
We present a spectrophotometric study of circumnuclear star forming regions (CNSFR) in the early type spiral galaxies: NGC 2903, NGC 3351 and NGC 3504, all of them of over solar metallicity according to standard empirical calibrations. A detailed determination of their abundances is made after careful subtraction of the very prominent underlying stellar absorption. It is found that most regions show the highest abundances in HII region-like objects. The relative N/O and S/O abundances are discussed. In is also shown that  CNSFR,  as a class, segregate from the disk HII region family, clustering around smaller ``softness parameter" -- $\eta$\' -- values, and therefore higher ionizing temperatures. 
\end{abstract}

\firstsection
\section{Introduction}
The inner parts of some spiral galaxies show higher star formation
rates than usual and this star formation is frequently arranged in a
ring or pseudo-ring pattern around their nuclei. 
In general, Circumnuclear Star
Forming Regions (CNSFR), also referred to as ``hotspots'', are alike
luminous and large disk HII regions, but look more compact and show
higher peak surphace brightness  (Kennicut et al. 1989). In many cases
they contribute substantially to the UV emission of the entire nuclear
region ({\it e.g.} Colina et al. 2002). Their H$\alpha$ luminosities overlap with those of HII galaxies being typically higher than 10$^{39}$ erg s$^{-1}$ which points to relatively massive ionizing star clusters. 
These regions are expected to show a high metallicity as corresponds to their position 
near the galactic bulge. They have considerable weight in the determination of abundance gradients, which in turn are widely used to constrain chemical evolution models, and constitute excellent
laboratories to study how star formation proceeds in high metallicity environments.

\section{Observations and reductions}
12 CNSFRs were observed with the 4.2m WHT at the Roque de los Muchachos Observatory
using the ISIS double spectrograph, with the EEV12 and TEK4 detectors in 
the blue and red arm respectively with the dichroic 
at $\lambda$7500 \AA\ .  Gratings  R300B in the blue
arm and R600R in the red arm were used, covering  the spectral ranges  $\lambda$3650
- $\lambda$7000 {\AA} in the blue and  ($\lambda$8850 -$\lambda$9650) in the near IR, yielding spectral  dispersions of 1.73 {\AA} pixel$^{-1}$ in the blue arm and  0.79 {\AA} pixel$^{-1}$ in the red arm.  A slit width of 1 arcesec was used.  
The nominal spatial sampling was 0.4 arcsec pixel$^{-1}$ in each frame and the average seeing for this night was {$\sim$}1.2 arcsec. 

%%%%%%%%%%%%%%%%%%%%%%%%%%%%%%%%%%%%%%%%%%%%%%%%%%%%%%%%%

\begin{figure}
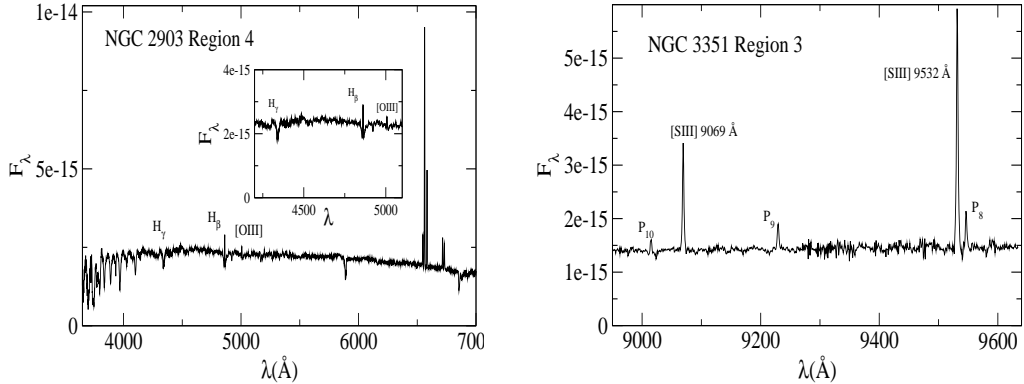

\includegraphics[width=2.55in,height=2.0in]{angelesdiazF1a.eps}
\hfill
\includegraphics[width=2.55in,height=2.0in]{angelesdiazF1b.eps}
\caption{{\it Left panel}: Blue spectrum of region 4 in NGC~2903  {\it Right panel}: Red spectrum of region 3 in NGC~3351.}
\label{fig}
\end{figure}

%%%%%%%%%%%%%%%%%%%%%%%%%%%%%%%%%%%%%%%%%%%%%%%%%%%%%%%%%

 The data were reduced using the IRAF (Image Reduction and Analysis Facility) package following standard procedures. The high spectral dispersion used in the near infrared allowed the almost complete elimination of the night-sky OH
emission lines and, in fact, the observed  $\lambda$9532/$\lambda$9069 ratio is close to the theoretical value of 2.48 in all cases. Telluric absorptions have been removed from the spectra 
of the regions by dividing by a relatively featureless continuum of a subdwarf star 
observed in the same night.

\section{Results and discussion}
Examples of blue and red spectra are shown in Figure 1.
Emission line fluxes were measured using the IRAF SPLOT software package.
The presence of a conspicuous underlying stellar population, more evident in the
blue spectra, in most observed regions complicates the measurements. An example of underlying absorption can be seen in the inset to the left panel of Figure 1. A two-component (emission and absorption) gaussian fit was performed in order to correct the Balmer lines for underlying absorption. An example of this procedure is shown in Figure 2.

%%%%%%%%%%%%%%%%%%%%%%%%%%%%%%%%%%%%%%%%%%%%%%%%%%%%%%%%%

\begin{figure}
\begin{center}
\includegraphics[height=1.5in]{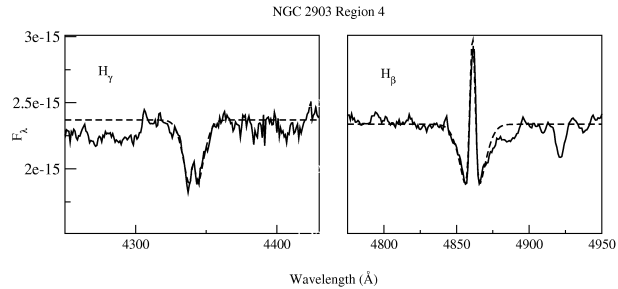}
\caption{Example of the fitting procedure used in order to correct the Balmer emission line intensities for the presence of the underlying stellar population.}
\label{fig}
\end{center}
\end{figure}

%%%%%%%%%%%%%%%%%%%%%%%%%%%%%%%%%%%%%%%%%%%%%%%%%%%%%%%%%

The low excitation of the regions, as evidenced by the weakness of the [OIII] $\lambda$ 5007 \AA\ line (see left panel of Figure 1), precludes the detection and measurement of the auroral [OIII] $\lambda$ 3463 \AA\ necessary for the derivation of the electron temperature. It is therefore impossible to obtain a direct determination of the oxygen abundances. Empirical calibrations have to be used instead. 
In the left panel of Figure 3 we show the calibration of oxygen abundance by the commonly used $O_{23}$ parameter defined as $([OII]\lambda 3727,29 + [OIII]\lambda 4959,5007)/H_\beta$ (Pagel et al. 1979). Data on HII galaxies, disc HII regions and CNSFR are shown. The HII region sample ( P\'erez-Montero \& D\'\i az 2005) has been divided in under-solar (open triangles)  and over-solar (filled triangles) according to the empirical criterion given by D\'\i az \& P\'erez-Montero (2000), {\it i. e.} $O_{23} \leq$ 0.47 and -0.5$\leq S_{23} \leq$ 0.28 \footnote{The sulphur abundance parameters $S_{23}$ is defined as  $([SII] + [SIII])/H_\beta$\\}. The HII galaxy data (filled squares) come from P\'erez-Montero \& D\'\i az (2003). CNSFR are represented by circles. Solid ones for our observed objects and open ones for regions in NGC~3310 and NGC~7714, known to have under-solar abundances (Pastoriza et. 1993; Gonz\'alez-Delgado et al. 1995). As can be seen the calibration is two-folded, shows a considerable scatter and its high abundance end is not well sampled. The position of the observed CNSFR is indicated. Their observed  $O_{23}$ values, lower than those of the lowest abundance galaxy known (IZw18), indicates that CNSFR belong to the high abundance branch of the calibration showing possibly the highest metallcities shown by HII region like objects. In the right panel of Figure 3 the position of the regions is indicated in the $N2$ ([NII]/ H$_\alpha$) abundance calibration diagram (Denicol\'o et al. 2002), which shows a linear behaviour. Again CNSFR appear to show the highest oxygen abundances.

%%%%%%%%%%%%%%%%%%%%%%%%%%%%%%%%%%%%%%%%%%%%%%%%%%%%%%%%%

\begin{figure}
\includegraphics[width=2.0in,height=2in]{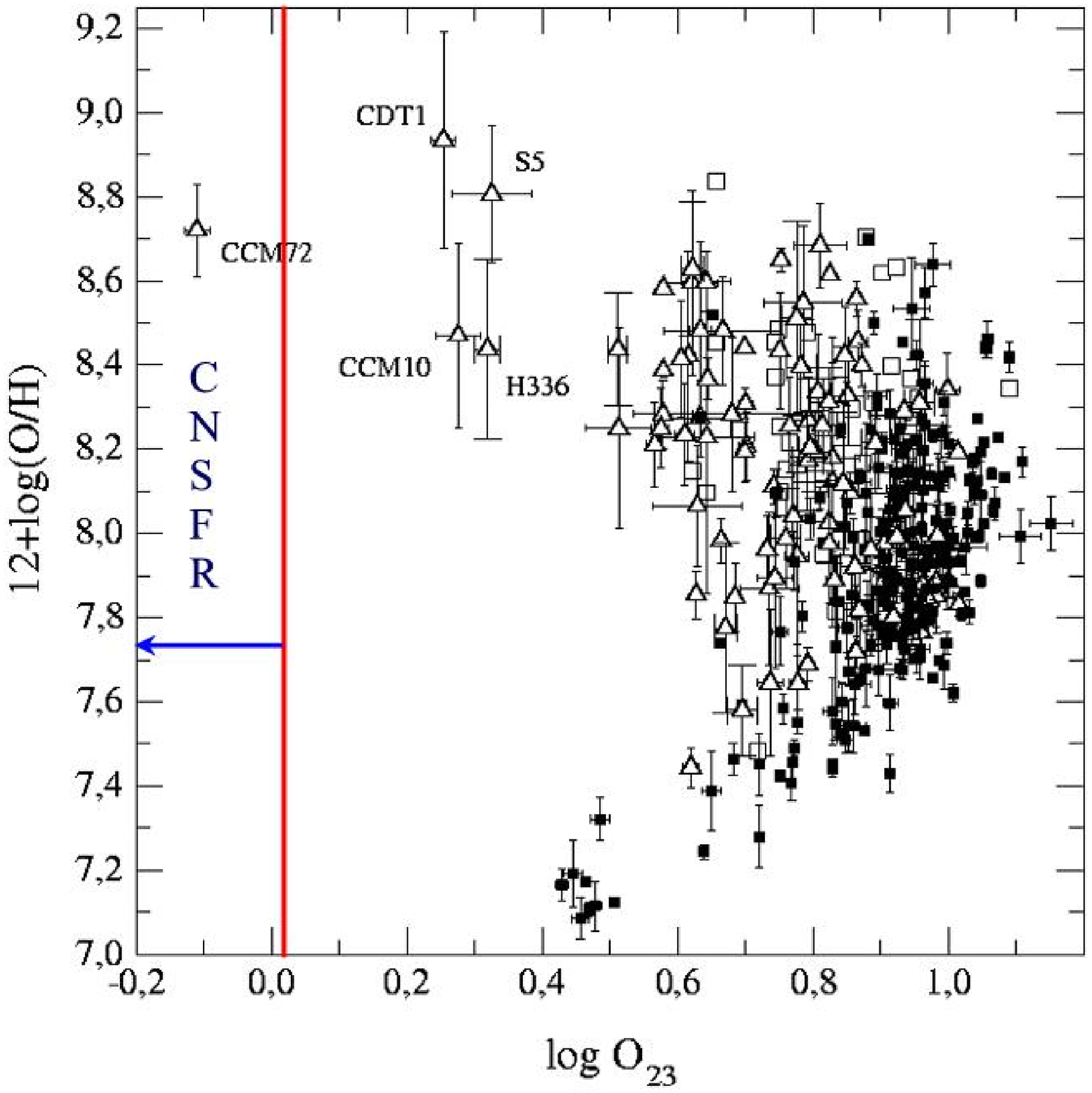}
\hfill
\includegraphics[width=2.0in,height=2in]{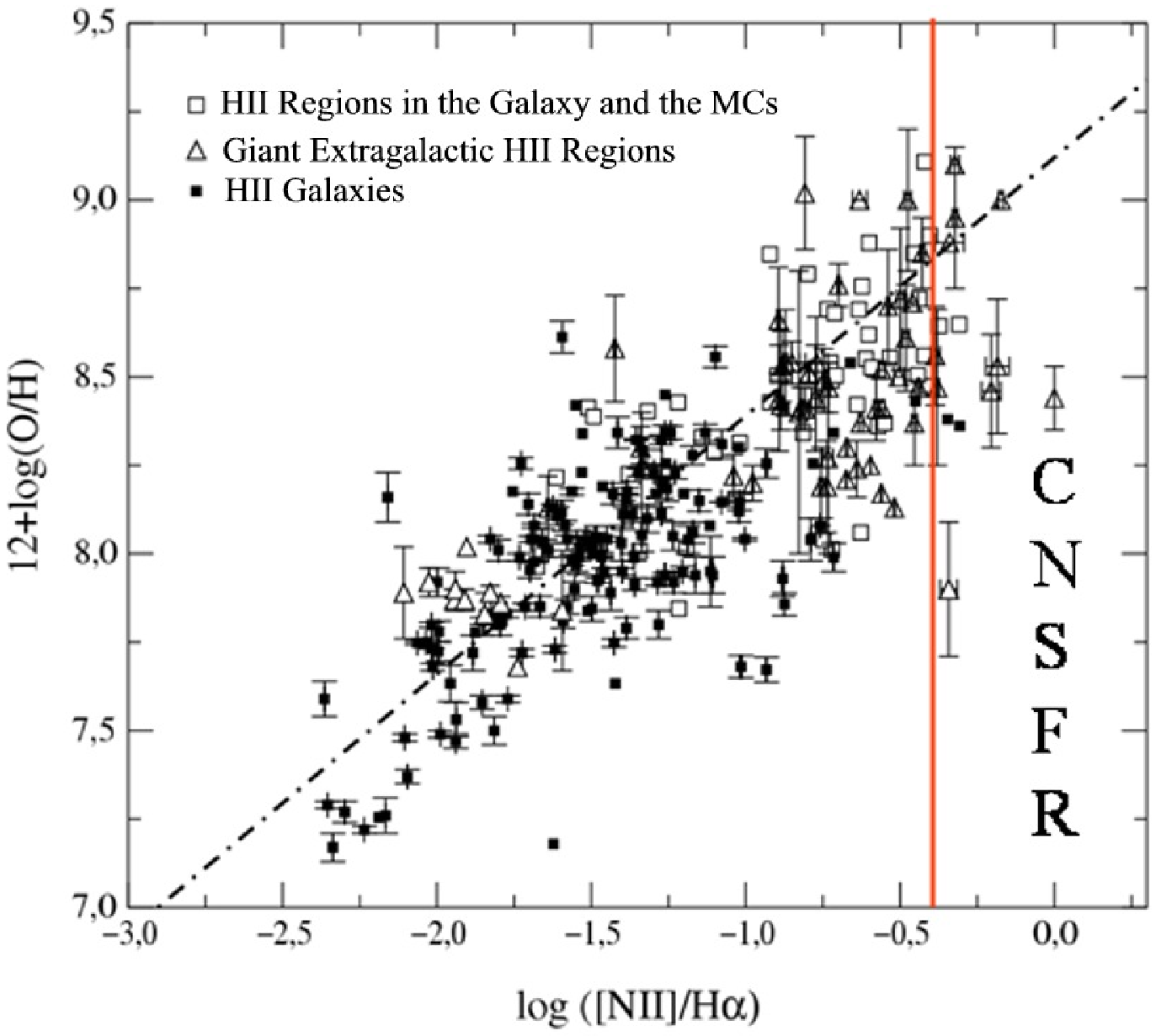}
\caption{{\it Left panel:} The $O_{23}$ abundance parameter calibration. {\it Right panel:} The $N2$ abundance parameter calibration. The location of CNSFR is indicated}
\label{fig}
\end{figure}

%%%%%%%%%%%%%%%%%%%%%%%%%%%%%%%%%%%%%%%%%%%%%%%%%%%%%%%%%

The left panel of Figure 4 shows the [NII]/[OII] ratio {\it versus} the $N2$ abundance parameter. Since a good correlation has been found to exist between the [NII]/[OII] ratio and the N$^+$/O$^+$ ionic abundance ratio, which in turn can be assumed to measure the N/O ratio, this graph is the observational equivalent of the N/O {\it vs} O/H diagram. We can see that a very tight correlation exists for all represented objects: HII galaxies, low and high abundance HII regions and CNSFR. Again our observed regions show the highest N/O ratios of the sample.

%%%%%%%%%%%%%%%%%%%%%%%%%%%%%%%%%%%%%%%%%%%%%%%%%%%%%%%%%

\begin{figure}
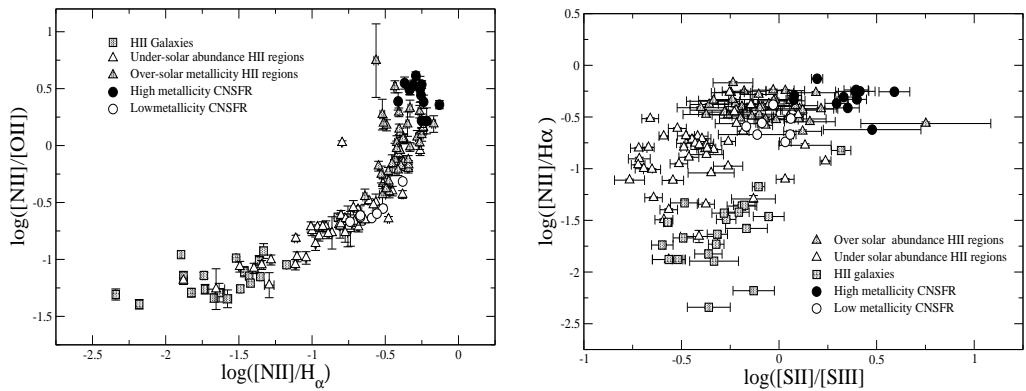

\includegraphics[width=2.55in,height=2.0in]{angelesdiazF4a.eps}
\hfill
\includegraphics[width=2.55in,height=2.0in]{angelesdiazF4b.eps}
\caption{{\it Left panel:} Empirical version of the N/O {\it vs} O/H relation.{\it Right panel:} The $N2$ abundance parameter as a funtion of excitation, as measured by the [SII]/[SIII] ratio.  \vspace*{0.5cm}}
\label{fig}
\end{figure}

%%%%%%%%%%%%%%%%%%%%%%%%%%%%%%%%%%%%%%%%%%%%%%%%%%%%%%%%%

The right panel of Figure 4 shows the run of the excitation degree with metallicity for the observed regions through the [SII]/[SIII] ratio, which has been shown to be a good ionization parameter indicator (D\'\i az et al. 1991), and the $N2$ parameter. It can be seen that the observed CNSFR show the lowest excitation of the sample.

A hint on the ionizing temperature of the regions can be obtained through the use of the $\eta$' parameter, which is a measure of the softness of the ionizing radiation (see V\'\i lchez \& Pagel 1988) and increases with decreasing ionizing temperature. The left panel of Figure 5  shows the run of $\eta$' with metallicity as parametrized by $N2$. Unexpectedly, CNSFR show values of $\eta$' higher than those of over-solar disc HII regions. This is better appreciated in the right panel of the figure where CNSFR are seen to segregate from disc HII regions in the [OII]/[OIII] {\it vs} [SII]/[SIII] diagram. The former cluster around the value of log$\eta$' = 0.0 (T${ion}\sim$ 40,000 K) while the latter cluster around log $\eta$' = 0.8 (T${ion}\sim$ 35,000 K).

This work has been supported by Spanish projects DGICYT- AYA-2004-08262-C03-03 and CM-ASTROCAM.

%%%%%%%%%%%%%%%%%%%%%%%%%%%%%%%%%%%%%%%%%%%%%%%%%%%%%%%%%

\begin{figure}
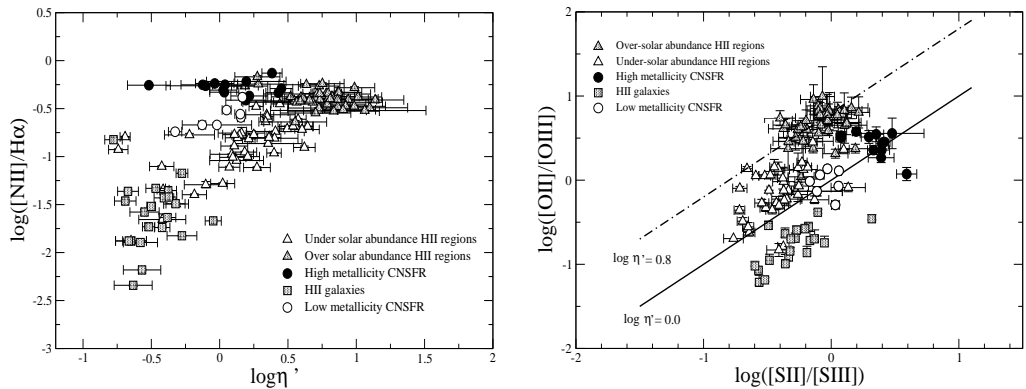

\includegraphics[width=2.55in,height=2.0in]{angelesdiazF5a.eps}
\hfill
\includegraphics[width=2.55in,height=2.0in]{angelesdiazF5b.eps}
\caption{{\it Left panel:} The $N2$ {\it vs} $\eta$' relation. {\it Right panel:} Logarithmic relation between the [OII]/[OIII] and [SII]/[SIII] line ratios}\label{fig}
\end{figure}

%%%%%%%%%%%%%%%%%%%%%%%%%%%%%%%%%%%%%%%%%%%%%%%%%%%%%%%%%

%%%%%%%%%%%%%%%%%%%%%%%%%%%%%%%%%%%%%%%%%%%%%%%%%%%%%%%%%%
%\end{document}

\end{document}